\documentclass[aps,twocolumn,floats,superscriptaddress,nopacs]{revtex4}
\usepackage{graphicx}
\usepackage{amssymb,amsmath}
\usepackage{color}
\usepackage{xcolor}
\usepackage{bm}
\usepackage{hyperref}
\begin{document}

\author{Aaram J. Kim}
\affiliation{Department of Physics, University of Fribourg, 1700 Fribourg Switzerland}
\affiliation{Department of Physics and Chemistry, DGIST, Daegu 42988, Korea}

\author{Evgeny Kozik}
\affiliation{Department of Physics, King's College London, Strand, London WC2R 2LS, United Kingdom}

\title{Misleading convergence of the skeleton diagrammatic technique: when the correct solution can be found}

\date{\today}
\begin{abstract}
Convergence of the self-consistent skeleton diagrammatic technique (SDT)---in which the full Green’s function is determined through summation of Feynman diagrams in terms of itself---to the wrong answer has been associated with the existence of non-perturbative branches of the Luttinger-Ward functional. Although it has been possible to detect misleading convergence without the knowledge of the exact result, the SDT has remained inapplicable in the regimes where this happens. We show that misleading convergence does not always preclude recovering the exact solution. In addition to the established mechanism, convergence of the SDT to the wrong answer can stem from divergence of the inherent diagrammatic series, which allows us to recover the exact solution by a modified SDT protocol based on controlled analytic continuation. We illustrate this approach by its application to the analytically solvable $(0+0)d$ Hubbard model, the Hubbard atom, and the $2d$ Hubbard model in a challenging strong-coupling regime, for which the SDT is solved with controlled accuracy by the diagrammatic Monte Carlo (DiagMC) technique. 
\end{abstract}

\maketitle

Green's functions are the universal language of quantum many-body theory~\citep{AGD} and the basis for state-of-the art computational methods of condensed matter, atomic, nuclear, and particle physics~\cite{Georges:1996zz,Maier2005,Rohringer2018,Gull2011,Blankenbecler1981,Zhang2003,Prokofev1998, Prokofev2007, VanHoucke2010,Kozik2010, VanHoucke2012}. In this framework, a property of a correlated system can be represented in a formally exact way by an infinite series of Feynman diagrams, comprising scattering events of the constituting particles to any order of the perturbation theory. The separation into the non-interacting particles and their coupling is, however, a largely arbitrary mathematical abstraction (see, e.g., Ref.~\cite{Kim2021homotopy} and references therein), and it is more natural to express Feynman diagrams in terms of the full Green's function $G$---the exact one-particle correlation function of the many-body system---which can then be determined self-consistently. This so-called skeleton (or bold-line) diagrammatic technique (SDT) is vastly more appealing both physically and mathematically: (i) $G$ describes observable quasiparticle properties and is experimentally measurable, (ii) as revealed by Baym and Kadanoff~\cite{Baym_Kadanoff}, expansions in terms of $G$ automatically respect conservation laws, (iii) they are also nonperturbative in a sense that each propagator in a diagram is renormalized to infinite order of the perturbation theory, (iv) the corresponding skeleton series thus contains fewer terms and (v), being built on the renormalized $G$ with the correct single-particle properties, the series is less prone to unphysical divergences and should generally converge faster, which makes the SDT \textit{a priori} more suitable for practical calculations. One catastrophic pitfall, however, can bring these benefits to naught: The bold-line series and the corresponding SDT can converge to the wrong unphysical answer~\cite{Kozik2015}. The misleading convergence is generally associated with multiple branches of the Luttinger-Ward functional (LWF)~\cite{Luttinger1960}, which underpins the bold-line expansions, making regimes belonging to a different branch inaccessible perturbatively~\cite{Kozik2015}. Connections of the problem to divergences of the irreducible vertex function, its physical origins and manifestations, and the very foundations of the many-body theory have become an active area of research~\cite{Schafer2013, Rossi:2015cx, Gunnarsson:2017c, Tarantino:2017dr,Thunstroem2018, Chalupa2018, KimReplica, Chalupa2021, VanHoucke2021,KimStrangeMetal, Adler2022}. Nonetheless, although practical means of detecting misleading convergence (without knowing the exact answer) have been developed~\cite{VanHoucke2021} and tested~\cite{KimStrangeMetal}, there is currently no general solution to the problem: The exact answer has remained inaccessible by the bold-line technique whenever the misleading convergence takes place.

Here we reveal a generic scenario of misleading convergence of the SDT, in which divergence of the underlying bold-line series plays a more fundamental role than the multivaluedness of the corresponding LWF. We demonstrate that a simple modification of the self-consistency protocol, which analytically continues the bold-line series beyond its convergence radius, in this case allows one to recover the exact solution behind the unphysical solution of the SDT. This approach enables controlled calculations in regimes where they have previously been deemed impossible. We illustrate this scenario by numerically exact results for the $(0+0)d$ Hubbard model and the Hubbard atom, which have remained the main testbeds for studies of the breakdown of many-body theories in the context of the multivaluedness of the LWF~\cite{Kozik2015,KimReplica,VanHoucke2021,KimStrangeMetal}. Finally, we apply our approach to obtain controlled results for the doped $2d$ Hubbard model in a challenging strong-coupling regime in the presence of misleading convergence. 

In the SDT, the Green's function $G$ is determined by consecutive approximations $G^{(n)}$, $G=\lim_{n \to \infty} G^{(n)}$, that self-consistently solve the Dyson equation
\begin{equation}
[G^{(n)}]^{-1}=G_0^{-1} - \Sigma^{(n)}[G^{(n)}, \xi=1], \label{eqn:Dyson}
\end{equation}
with $G_0$ being the noninteracting Green's function. Here, the self-energy functional $\Sigma^{(n)}[\mathcal{G}, \xi]$ (we use the symbol $\mathcal{G}$ for the arbitrary argument) is computed as the partial sum of all bold-line Feynman diagrams constructed from a given Green's function $\mathcal{G}$ up to the order $n$ in the powers of coupling~\cite{AGD},
\begin{equation}
\Sigma^{(n)}[\mathcal{G}, \xi]=\sum_{m=1}^{n} a_m[\mathcal{G}] \xi^m, \label{eqn:series}
\end{equation}
while the series coefficients $a_m[\mathcal{G}]$ depend only on the function $\mathcal{G}$, and $\xi$ is a formal expansion parameter associated with each interaction line and set to $\xi=1$ in final expressions; we denote $\Sigma^{(n)}[\mathcal{G}, \xi=1]$ by $\Sigma^{(n)}[\mathcal{G}]$. It has recently been realized~\cite{Kozik2015, Rossi:2015cx,VanHoucke2021, Gunnarsson:2017c, KimReplica} that the functional $\Sigma[\mathcal{G}]$ and the underpinning LWF $\Phi[\mathcal{G}]$, $\delta \Phi/ \delta \mathcal{G}=\Sigma$, may have at least two branches with serious consequences for the SDT. The bold-line expansion (\ref{eqn:series}) defines the (``weak-coupling'') branch $\Sigma^w[\mathcal{G}]$ that is analytically connected to the $\xi \to 0$ limit, $\Sigma^w[\mathcal{G}] = \lim_{n \to \infty} \Sigma^{(n)}[\mathcal{G}]$. However, the exact solution $G_\mathrm{exact}$ can happen to belong to another (``strong-coupling'') branch $\Sigma^s[\mathcal{G}]$, which is not accessible by the skeleton expansion (\ref{eqn:series}). We can thus define two mutually exclusive (depending on the system parameters) regimes: the ``perturbative''~\footnote{Note that the natural notations ``perturbative'' and ``weak-coupling'' do not imply weakly-correlated, as highly non-trivial physics can emerge in the $\Sigma^w$ branch, while the series (\ref{eqn:series}), built on the nonperturbatively renormalized $G$, could converge only at very high orders.},
\begin{equation}
G_\mathrm{exact}  = \big[G_0^{-1} - \Sigma^w [G_\mathrm{exact}] \big]^{-1} \;\;\; \text{(P)}, \label{eqn:perturbative_regime}
\end{equation}
and ``nonperturbative'',
\begin{equation}
G_\mathrm{exact}  = \big[G_0^{-1} - \Sigma^s [G_\mathrm{exact}] \big]^{-1}\;\;\; \text{(NP)}.\label{eqn:nonperturbative_regime}
\end{equation}
Here $G_\mathrm{exact}$ is generally a function of momentum $\mathbf{k}$ and Matsubara frequency $\omega_m$ (suppressed for clarity), so that the switchover from P to NP regime, e.g., with increasing the coupling strength $U$ or lowering the temperature $T$, happens for each $(\mathbf{k}, \omega_m)$ individually, typically first at the lowest Matsubara frequency followed sequentially by the others~\cite{Kozik2015, KimReplica}.
In a typical scenario discussed previously~\cite{Kozik2015, Rossi:2015cx, VanHoucke2021}, starting with $G_\mathrm{exact}$ on the $\Sigma^w$ branch (\ref{eqn:perturbative_regime}), a continuous evolution of $\Sigma^{w,s}[G_\mathrm{exact}]$ with system parameters leads to their crossing,  $\Sigma^w[G_\mathrm{exact}]=\Sigma^s[G_\mathrm{exact}]$, beyond which $G_\mathrm{exact}$ switches to the $\Sigma^s$ branch (\ref{eqn:nonperturbative_regime})~\cite{Kozik2015, Rossi:2015cx, VanHoucke2021}.
 This degeneracy of the LWF thus marks the P-to-NP switchover by divergence of the corresponding irreducible vertex function $\Gamma=\delta \Sigma[G_\mathrm{exact}]/\delta \mathcal{G}$, discovered and studied in detail in diverse (momentum-independent) systems~\cite{Schafer2013, Rossi:2015cx, Gunnarsson:2017c, Tarantino:2017dr,Thunstroem2018, Chalupa2018, Chalupa2021, Adler2022}.
 More generally, $\Sigma^s[G_\mathrm{exact}]$ and $\Sigma^w[G_\mathrm{exact}]$ can swap discontinuously across the P-to-NP switchover, while $G_\mathrm{exact}$ evolves smoothly without the vertex function divergence, as, e.g., in the $(0+0)d$ Hubbard model for $\mathrm{Re}G_0 \neq 0$, which mimics a doped (from half-filling) lattice system.

Clearly, in the NP regime the SDT (\ref{eqn:Dyson}), (\ref{eqn:series}) can not find $G_\mathrm{exact}$. If it does converge at all, the obtained solution must necessarily be incorrect and referred to as ``unphysical'', $G_\mathrm{unphys}$. This is a catastrophic problem for unbiased approaches based on explicit evaluation of the SDT such as diagrammatic Monte Carlo (DiagMC)~\cite{Prokofev1998,Prokofev2007, Prokofev2008, VanHoucke2010, Kozik2010}, which makes control of accuracy in bold-line calculations meaningless if the convergence to $G_\mathrm{exact}$ cannot be guaranteed. Worse yet, building on the correct $G_0$ via the Dyson Eq.~(\ref{eqn:Dyson}), $G_\mathrm{unphys}$ generically does not exhibit clearly unphysical features~\cite{Kozik2015,VanHoucke2021,KimStrangeMetal}, so even detecting the misleading convergence without knowing $G_\mathrm{exact}$ in advance is a serious fundamental problem. 

Thankfully, a \textit{necessary} condition for the misleading convergence of the SDT has been developed~\cite{Rossi_shifted_action_2016, VanHoucke2021}: In the NP regime, convergence is only possible if the self-consistency (\ref{eqn:Dyson}) fine-tunes the solution $G\equiv G_\mathrm{unphys}$ so that the series $\Sigma^{(n)}[G_\mathrm{unphys}]$ is exactly at its convergence radius. More specifically, if [for each $(\mathbf{k}, \omega_m)$] we define the location of the singularity $\xi_s$ closest to the origin in the complex $\xi$-plane, e.g., via the ratio test 
\begin{equation}
\xi_s[\mathcal{G}]=\lim_{m \to \infty} a_{m-1}[\mathcal{G}]/a_{m}[\mathcal{G}], \label{eq:ratio_test}
\end{equation}
then we must obtain $|\xi_s[G_\mathrm{unphys}]|=1$ in the NP regime at least for one momentum-frequency value, and we will show that this condition is actually stricter: $\xi_s[G_\mathrm{unphys}]=1$.
Our central observation is that the same scenario of misleading convergence takes place in the \textit{a priori} tractable P regime (\ref{eqn:perturbative_regime}) whenever $\Sigma^{(n)}[G_\mathrm{exact}]$ happens to be divergent due to a singularity with $|\xi_s[G_\mathrm{exact}]|<1$. 


\begin{figure}[t!]
\centering
\includegraphics[width=1.0\columnwidth]{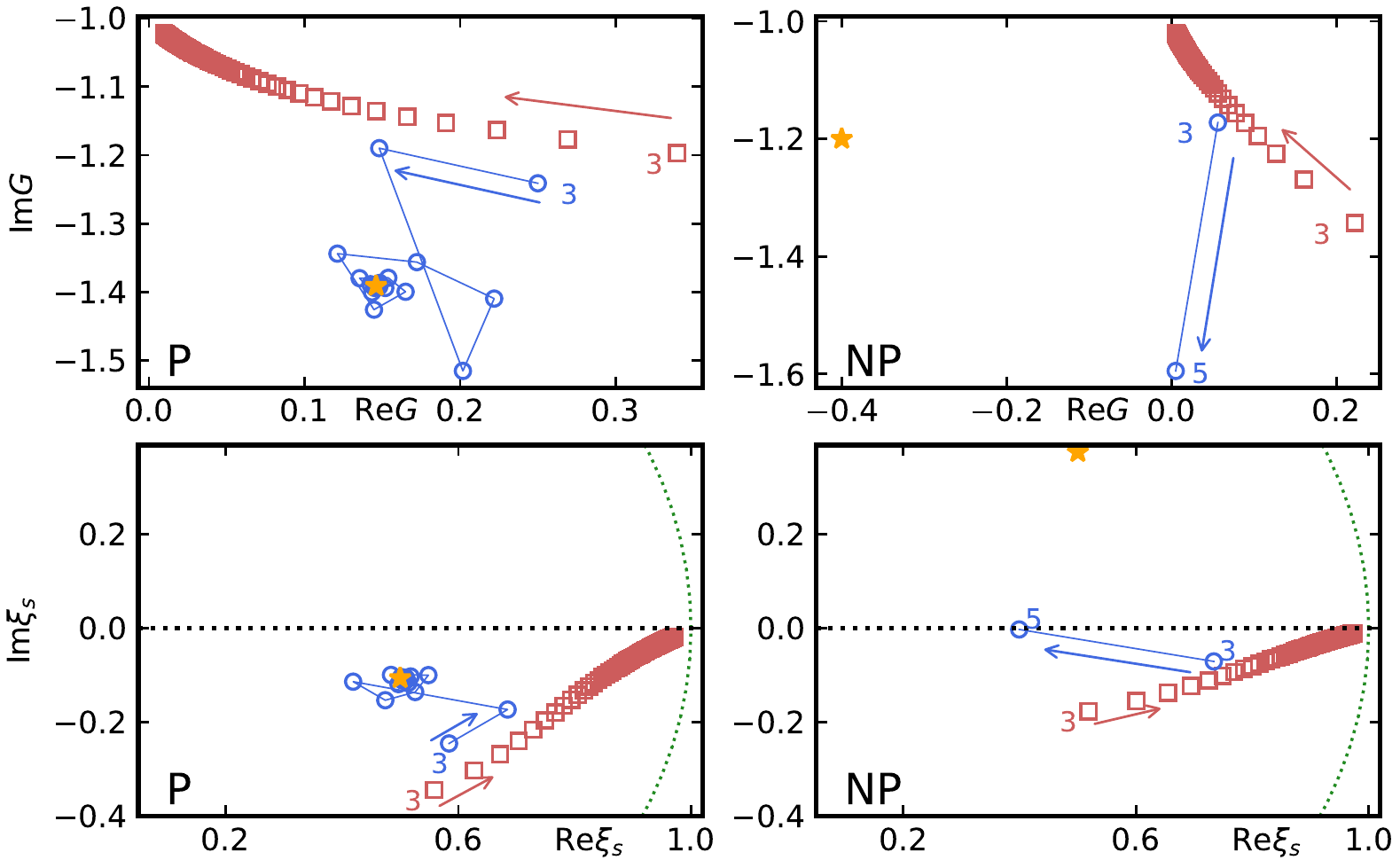}
\caption{Bold-line-diagrammatic solution of the $(0+0)d$ Hubbard model (\ref{eqn:0+0_action}) at $U=1/4$ for $G_0=1.8 e^{-i\pi/4}$ (\textbf{left column}) and $G_0=2\sqrt{2}e^{-i\pi/4}$ (\textbf{right column}), which are in the P (\ref{eqn:perturbative_regime}) and NP (\ref{eqn:nonperturbative_regime}) regimes, respectively. \textbf{Top row}: the evolution of $G^{(n)}$ with the diagram order $n$ obtained by the SDT (\ref{eqn:Dyson}), (\ref{eqn:series}) (red squares) and the mSDT (\ref{eqn:series}), (\ref{eqn:Dyson-modified}) (blue circles), the direction of increasing $n$ is shown by the arrows, and the exact solution $G_\mathrm{exact}$ is marked by the star. \textbf{Bottom row}: The corresponding location of the singularity $\xi_s[G^{(n)}]$ obtained by Eq.~(\ref{eq:ratio_test}); $\xi_s[G_\mathrm{exact}]$ is marked by the star. Note that for both values of $G_0$ the SDT slowly converges to the same $G_\mathrm{unphys}=-i$ with $\xi_s[G_\mathrm{unphys}]=1$ in the $n \to \infty$ limit, while the mSDT recovers $G_\mathrm{exact}$ in the P regime already for $n \gtrsim 8$. 
}
\label{fig:0+0d_SDT}
\end{figure}

The mechanism can be illustrated by the $(0+0)d$ Hubbard model~\cite{Rossi:2015cx,VanHoucke2021, Gunnarsson:2017c, KimReplica} for Grassmann variables (as opposed to fields) $\psi_\sigma$, with $\sigma$ being the spin index, the SDT for which can be constructed analytically: The self-energy becomes a \textit{function} of the complex number $\mathcal{G}$, while the exact solution $G_\mathrm{exact}=-\langle \psi_\sigma \overline{\psi}_\sigma \rangle$ is found through averaging with the effective action
\begin{equation}
	\mathcal{S} = -\sum^{}_{\sigma=\uparrow, \downarrow}\overline{\psi}_\sigma G_{0}^{-1}\psi_\sigma + U n_{\uparrow}n_{\downarrow}~,
	\label{eqn:0+0_action}
\end{equation}
where the parameters $G_0$ and $U$ are generally complex numbers and $n_{\sigma}=\overline{\psi}_\sigma \psi_\sigma$. Action~(\ref{eqn:0+0_action}) has been introduced~\cite{Rossi:2015cx,VanHoucke2021} as a toy model exemplifying the structure of the LWF to study the misleading convergence 
when $G_0$ is purely imaginary and $U$ is a positive real number~~\cite{KimReplica}, mimicking the half-filled Hubbard model. It was further noted~\cite{KimReplica} that a general Hubbard model for the fermionic Grassmann fields $\psi_{\sigma \mathbf{k} \omega_m}$ at a fixed $\mathbf{k}$ and $\omega_m$ can be reduced to the model (\ref{eqn:0+0_action}) with some effective momentum-frequency-dependent $G_0$ and $U$. Thus, the general $(0+0)d$ action underpins the structure of the many-body theory in the full Hubbard model and deserves further investigation.

The two branches in the model (\ref{eqn:0+0_action}) are given by $\Sigma^{w,s}[\mathcal{G}] = 2U\mathcal{G}/[1\pm \sqrt{1+4 U\mathcal{G}^2}]$, while $G_\mathrm{exact}= G_0^{-1}/[G_0^{-2} - U]$, so that the P and NP regimes can be explicitly identified, and the Taylor expansion of $\Sigma^{w}[\mathcal{G}, \xi]=2\xi U\mathcal{G}/[1 + \sqrt{1+4\xi U\mathcal{G}^2}]$ about $\xi=0$ is the bold-line diagrammatic series (\ref{eqn:series}). In Fig.~\ref{fig:0+0d_SDT}, the SDT is applied to both P (for $G_0=1.8 e^{-i\pi/4}$) and NP (for $G_0=2\sqrt{2}e^{-i\pi/4}$) regimes at $U=1/4$. Surprisingly, despite the different non-interacting $G_0$, the solution of the SDT in the $n \to \infty$ limit is the same $G_\mathrm{unphys}=-i$ (top row of Fig.~\ref{fig:0+0d_SDT}), corresponding to $\xi_s[G_\mathrm{unphys}]=1$ (bottom row of Fig.~\ref{fig:0+0d_SDT}). Thus, the necessary condition of misleading convergence~\cite{Rossi_shifted_action_2016, VanHoucke2021} $|\xi_s|=1$ is satisfied and the SDT is inapplicable in both cases. The exact solution cannot be obtained for $G_0=1.8 e^{-i\pi/4}$ in the P regime, which should in principle be tractable by the SDT, because the diagrammatic series $\Sigma^{(n)}[G_\mathrm{exact}]$ turns out to be divergent 
and thus the self-consistency (\ref{eqn:Dyson}) cannot be satisfied by $G_\mathrm{exact}$, albeit for a fundamentally different reason in the NP regime. Instead, if there is a value $G$ such that $\xi_s[G]=1$, it asymptotically satisfies Eq.~(\ref{eqn:Dyson}) in the $n \to \infty$ limit \textit{for any} $G_0$, since an infinitesimal detuning from $G$ allows $\Sigma^{(n \to \infty)}[G]$ to take any value. Thus, if Eq.~(\ref{eqn:Dyson}) cannot be satisfied by $G_\mathrm{exact}$---be it for $G_\mathrm{exact}$ belonging to the strong-coupling branch (\ref{eqn:nonperturbative_regime}) or divergence of $\Sigma^{(n)}[G_\mathrm{exact}]$ with $\xi_s<1$---the SDT generically finds $G_\mathrm{unphys}$ with $\xi_s[G_\mathrm{unphys}]=1$. Note that the condition $\xi_s=1$ is essential also because singularities elsewhere in the complex plane (except, perhaps, on the real axis with $0< \xi_s < 1$) can be avoided altogether by reformulating the problem identically in terms of the homotopic action~\cite{Kim2021homotopy}.

This mechanism suggests that the misleading convergence of the SDT does not always preclude recovering the exact answer. Indeed, if $G_\mathrm{exact}$ is in the P regime and the misleading convergence is caused by the divergence of the corresponding bold-line series with $|\xi_s[G_\mathrm{exact}]|<1$ then the $\Sigma^w$ branch can be reconstructed by an appropriate analytic continuation~\cite{janke1998resummation, baker1961Dlog, Hunter1979} of the divergent series $\Sigma^{(n)}$ beyond the singularity $\xi_s$. To this end, we can modify the SDT (\ref{eqn:Dyson}), (\ref{eqn:series}) replacing the partial sum $\Sigma^{(n)}$ in the Dyson Eq.~(\ref{eqn:Dyson}) with its analytic continuation $\tilde{\Sigma}^w$,
\begin{equation}
[G^{(n)}]^{-1}=G_0^{-1} - \tilde{\Sigma}^w\{ a_m[G^{(n)}] \}, \label{eqn:Dyson-modified}
\end{equation}
where the notation $\tilde{\Sigma}^w\{ a_m[G^{(n)}] \}$ emphasises that the analytic continuation is constructed from the set of series coefficients $\{a_m[G^{(n)}]\}$, $m=1, \ldots, n$, comprising the original partial sum (\ref{eqn:series}). Clearly, when the series (\ref{eqn:series}) converges, the modified SDT (mSDT) (\ref{eqn:series}), (\ref{eqn:Dyson-modified}) is identical to the original one, but it allows one to recover in the $n \to \infty$ limit the solution (\ref{eqn:perturbative_regime}) even if $\Sigma^{(n)}[G_\mathrm{exact}]$ diverges. 

Here and throughout we use the Dlog-Pad\'e method~\cite{baker1961Dlog, Hunter1979} to construct $\tilde{\Sigma}^w$, which is based on associating the coefficients $\{a_m\}$ truncated at the order $n$ with the Taylor series of an ansatz function with power-law singularities. This approach was shown to enable analytic continuation with controllable error bars in the general case of the full Hubbard model~\cite{Simkovic2019, Fedor2020, Kim_PRL_2020}. The Dlog-Pad\'e value of $\tilde{\Sigma}^w$ is guaranteed to asymptotically approach $\Sigma^w$ in the $n \to \infty$ limit, and thus applying Eq.~(\ref{eqn:Dyson-modified}) is not different from the original technique (\ref{eqn:Dyson}) when the series converges.

Figure~\ref{fig:0+0d_SDT} (left column) demonstrates that the mSDT (\ref{eqn:series}), (\ref{eqn:Dyson-modified}) is able to find the exact solution $G_\mathrm{exact}$ in the P regime despite the misleading convergence of the SDT. Note the quick convergence to $G_\mathrm{exact}$ already at $n \gtrsim 8$, in contrast to the extremely slow convergence (to $G_\mathrm{unphys}$) of the SDT, which could be used as a circumstantial diagnostic of misleading convergence.
In the NP regime, in contrast (Fig.~\ref{fig:0+0d_SDT} right column), the mSDT does not show convergence beyond $n=5$ and up to $n \sim 200$ considered. This is because the $\Sigma^w$ branch recovered by the analytic continuation is inconsistent with $G_\mathrm{exact}$ given by Eq.~(\ref{eqn:nonperturbative_regime}). Thus, any nontrivial solution of the mSDT in this regime must be a result of evaluating the analytic continuation $\tilde{\Sigma}^w[\mathcal{G}]$ beyond its validity domain, where it can be different from the exact $\Sigma^w[\mathcal{G}]$. Indeed, for the highest order $n=5$ at which the mSDT finds a solution $G$, the (nearest to the origin)
singularity $\xi_s[G]$ in $\tilde{\Sigma}^w[G]$ is located on the real axis with $0<\xi_s<1$ (Fig.~\ref{fig:0+0d_SDT} bottom right), meaning that the approximant $\tilde{\Sigma}^w [G,\xi=1]$ is evaluated on its branch cut.

\begin{figure}[t!]
\centering
\includegraphics[width=1.0\columnwidth]{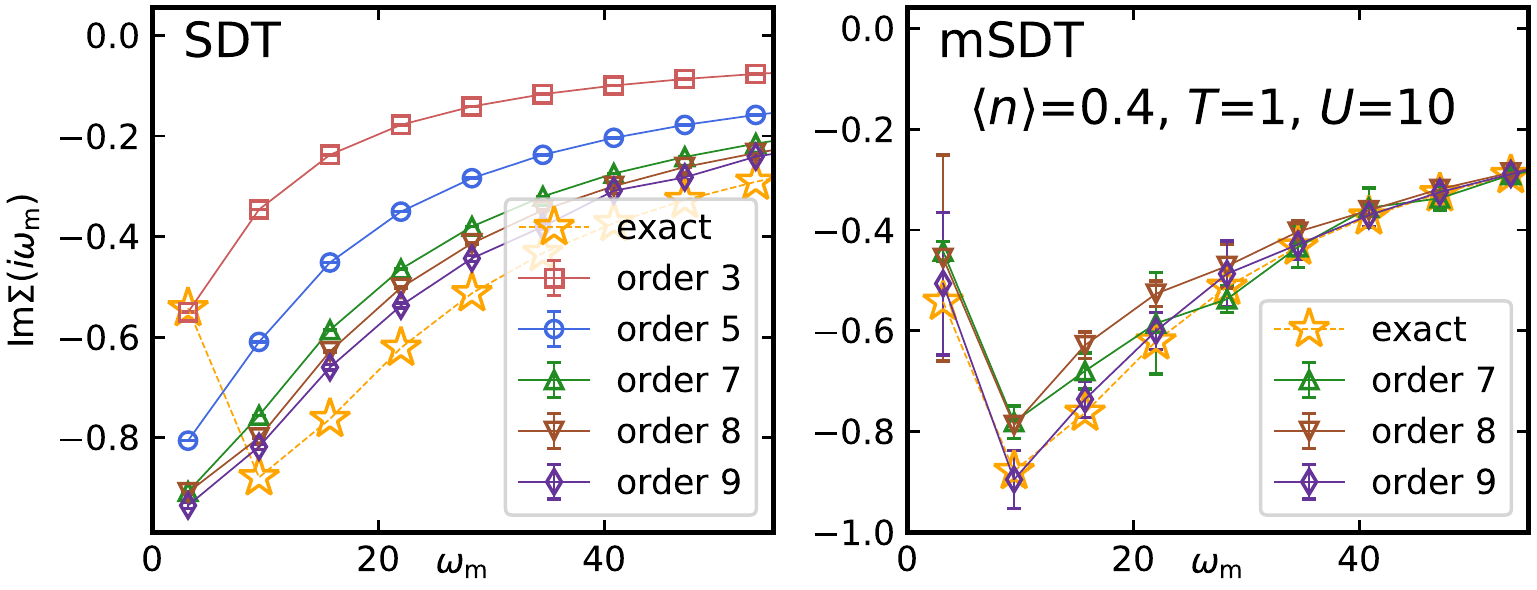}
\caption{Hubbard atom in the P regime [$U=10$, $T=1$, and density $\langle n \rangle=0.4$]: (\textbf{Left}) Misleading convergence of the SDT (\ref{eqn:Dyson}), (\ref{eqn:series}): The partial sums of the self-energy $\mathrm{Im} \Sigma^
{(n)}(i\omega_m) [G^{(n)}]$, plotted as functions of the Matsubara frequency $\omega_m$, approach an unphysical solution at large diagram orders $n$ -- the (very different) exact solution is shown by the stars. (\textbf{Right}) Recovery of the exact solution by the the mSDT (\ref{eqn:series}), (\ref{eqn:Dyson-modified}): the controlled analytic continuation of the self-energy $\mathrm{Im} \tilde{\Sigma}^
{w}(i\omega_m)[G^{(n)}]$ obtained by the mSDT (\ref{eqn:series}), (\ref{eqn:Dyson-modified}) finds the exact solution for large $n$.}
\label{fig:HA_P_modified_SDT}
\end{figure}
%
\begin{figure}[t!]
\centering
\includegraphics[width=1.00\columnwidth]{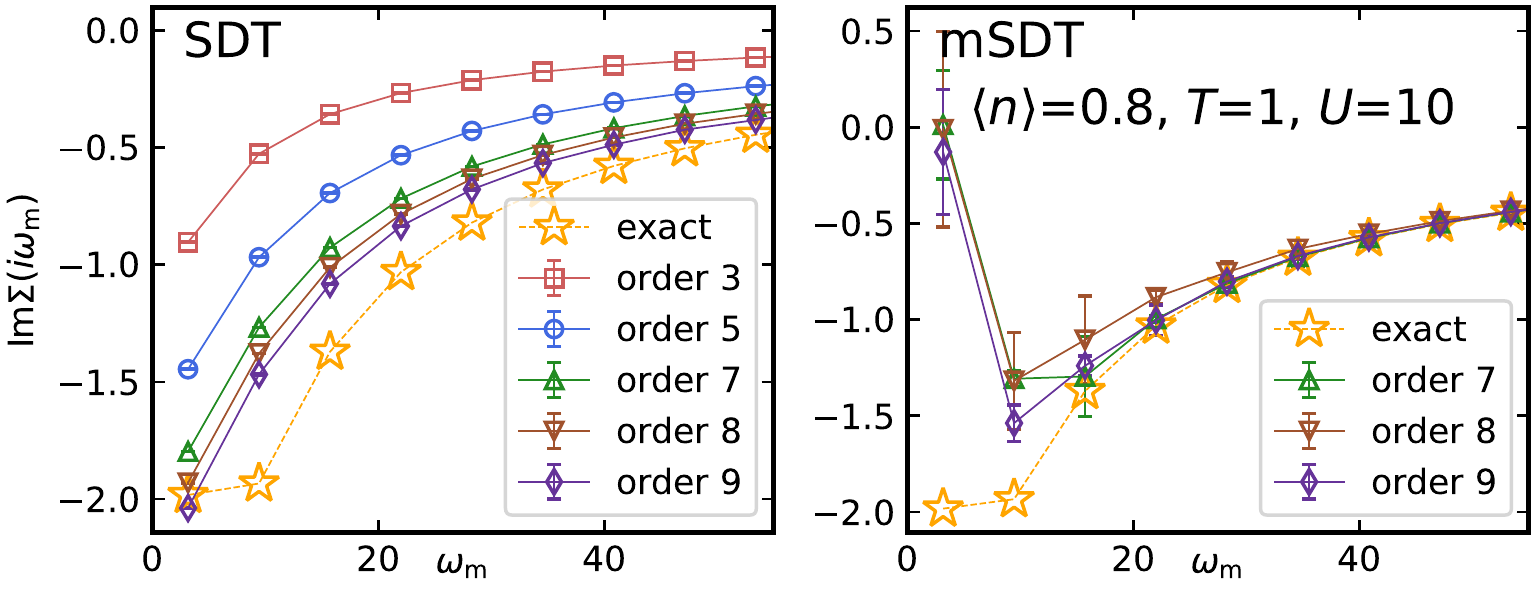}
\caption{Hubbard atom in the NP regime [$U=10$, $T=1$, and density $\langle n \rangle=0.8$]: (same notations as in Fig.~\ref{fig:HA_P_modified_SDT}) Both the SDT and mSDT approaches fail to find the exact solution. }
\label{fig:HA_NP_modified_SDT}
\end{figure}

The more realistic case of the Hubbard atom, in which only spatial fluctuations are neglected, has been instrumental in discovering and analyzing consequences of the multivaluedness of the LWF~\cite{Schafer2013, Kozik2015, Gunnarsson:2017c,Chalupa2018,Thunstroem2018, Chalupa2021, VanHoucke2021, Adler2022}. With regard to misleading convergence, it exhibits a qualitatively similar behavior to the full Hubbard model~\cite{Kozik2015}, but is more instructive due to the absence of momentum dependence and availability of the exact analytic solution $G_\mathrm{exact}$. It is a realistic test case for the SDT because---just as in the full Hubbard model---we have to resort to the DiagMC technique~\cite{VanHoucke2010, Kozik2010} for numerically exact evaluation of the self-energy series (\ref{eqn:series}) to order $n \sim 9$, 
with the only error being the statistical uncertainty. In Fig. \ref{fig:HA_P_modified_SDT}, both the SDT (\ref{eqn:Dyson}), (\ref{eqn:series}) and mSDT (\ref{eqn:series}), (\ref{eqn:Dyson-modified}) techniques are applied to the Hubbard atom at $U=10$, temperature $T=1$ (throughout we use the units of hopping $t=1$ of the full Hubbard model), and density $\langle n \rangle=0.4$. It is seen that the partial sums of the self-energy $\mathrm{Im} \Sigma^
{(n)}(i\omega_m) [G^{(n)}]$ found by the SDT (left panel) approach at large orders $n$ a solution that is markedly different from the exact result shown by the stars. The wrong self-energy does not exhibit any obviously unphysical features and can not be immediately ruled out without knowing the exact solution. However, when the mSDT is applied to the same system (right panel), the resulting analytic continuation of the self-energy $\mathrm{Im} \tilde{\Sigma}^
{w}(i\omega_m)[G^{(n)}]$ is seen to converge to the exact solution. For comparison, Fig.~\ref{fig:HA_NP_modified_SDT} illustrates the application of the SDT and mSDT to the Hubbard atom in the NP regime ($U=10$, $T=1$, and density $\langle n \rangle=0.8$). As expected, in this case both techniques fail to recover the exact answer.

\begin{figure}[t!]
\centering
\includegraphics[width=1.00\columnwidth]{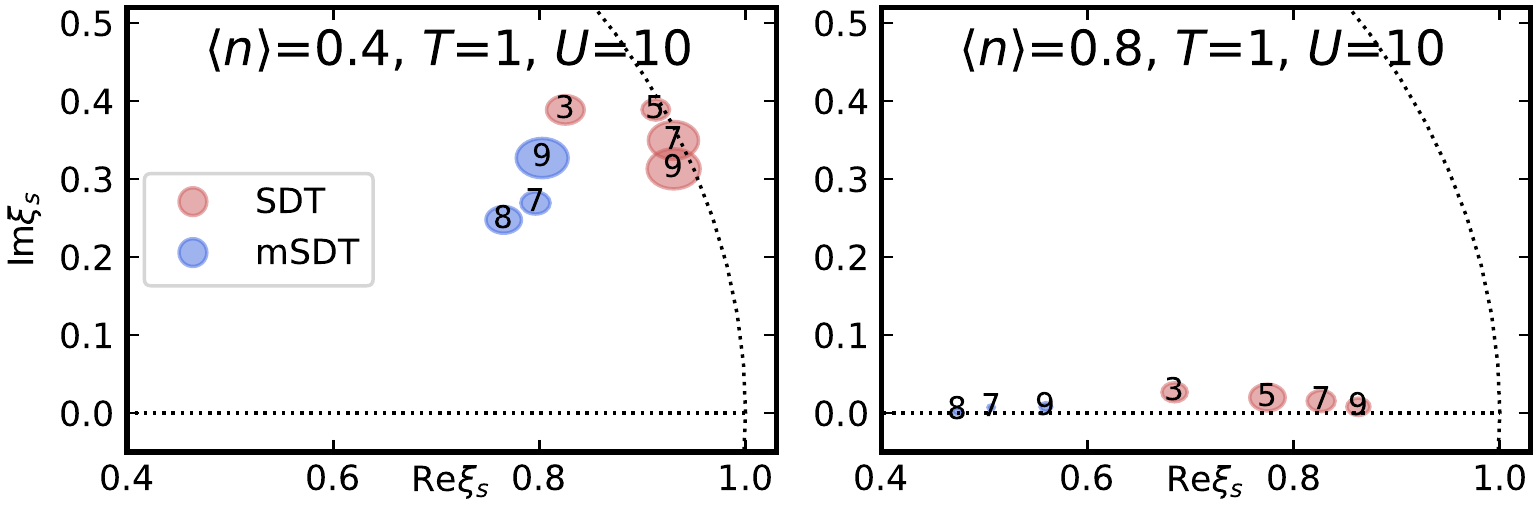}
\caption{Location of the singularity $\xi_s[G^{(n)}]$ for the SDT (red circles) and mSDT (blue circles) solutions $G^{(n)}$ of the Hubbard atom found in Fig.~\ref{fig:HA_P_modified_SDT} (P regime, \textbf{left panel}) and Fig.~\ref{fig:HA_NP_modified_SDT} (NP regime, \textbf{right panel}), with the diagram orders $n$ labelled by the numbers. The circle sizes represent the error bars of $\xi_s$.}
\label{fig:HA_singularities}
\end{figure}

For practical calculations, it is crucial to be able to diagnose the misleading convergence of the SDT and mSDT without the \textit{a priori} knowledge of the exact solution. To this end, in Fig.~\ref{fig:HA_singularities} we plot the estimates of the singularities $\xi_s[G^{(n)}]$ for the solutions $G^{(n)}$ found in Fig.~\ref{fig:HA_P_modified_SDT} (P regime, left panel) and Fig.~\ref{fig:HA_NP_modified_SDT} (NP regime, right panel). As in the $(0+0)d$ case, $\xi_s[G^{(n)}]$ for the SDT solution in both P and NP regimes drifts very slowly but consistently with increasing $n$ (marked by the numbers) towards $\xi_s=1$ from within the unit circle~\footnote{Curiously, similarly to the lower row of Fig.~\ref{fig:0+0d_SDT}, $\xi_s$ has a smaller imaginary part in the NP regime, albeit to a much greater extent.}. This behavior implies that the SDT solution cannot be trusted. The failure of the mSDT in the NP regime (Fig.~\ref{fig:HA_singularities} right) is evident from the fact that the corresponding singularities are all on the real axis within the error bars and $0<\xi_s[G^{(n)}]<1$, i.e. $\tilde{\Sigma}^w[G^{(n)}]$ is evaluated on its branch cut. In the P regime (Fig.~\ref{fig:HA_singularities} left), in contrast, the mSDT singularities are clearly away from the real axis and, although there are some deviations between the estimates for different orders comparable to the error bars, the data are consistent with $|\xi_s[G_\mathrm{exact}]|<1$. This comprises the criterion of validity of the mSDT solution.

\begin{figure*}[ht!]
\centering
\includegraphics[width=0.95\textwidth]{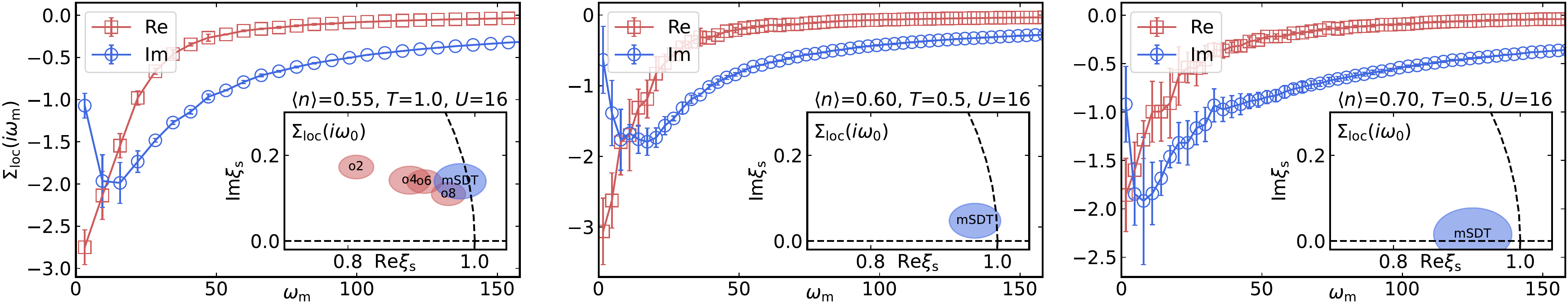}
\caption{The converged mSDT solution of the $2d$ Hubbard model at $U=16t$ and different densities and temperatures (from left to right): $(\langle n \rangle, T)=(0.55, 1.0t), (0.6, 0.5t), (0.7, 0.5t)$ . The local self-energy $\Sigma_\mathrm{loc}(i\omega_m)=\int \frac{d \mathbf{k}}{(2\pi)^2} \Sigma(\mathbf{k}, i\omega_m)$ is shown as a function of the Matsubara frequency. The insets are the loci of the singularities $\xi_s$ in the corresponding diagrammatic series at the lowest Matsubara frequency, with the circle size corresponding to the conservative error bar of $\xi_s$. In the left panel, $\xi_s[G^{(n)}]$ for the SDT solution are shown in the notations of Fig.~\ref{fig:HA_singularities}.}
\label{fig:HM_data}
\end{figure*}

Finally, we apply this approach to the doped $2d$ Hubbard model at $U=16$, where no controlled benchmark by any other method is available. Fig.~\ref{fig:HM_data} shows the converged with diagram order local self-energy $\Sigma_\mathrm{loc}(i\omega_m)=\int \frac{d \mathbf{k}}{(2\pi)^2} \Sigma(\mathbf{k}, i\omega_m)$ obtained by the mSDT for different densities and temperatures. At these parameters the SDT exhibits misleading convergence, as illustrated by the underlying singularity structure in the inset for $\langle n \rangle=0.55$, $T=1.0$. The singularities corresponding to the converged mSDT solutions for $\langle n \rangle=0.55$, $T=1.0$ and $\langle n \rangle=0.6$, $T=0.5$ shown in the insets, pass the \textit{a priori} validity test. However, for $T=0.5$, $\langle n \rangle=0.7$, $\xi_s$ is consistent, given the error bar, with $\mathrm{Im}\xi_s=0$, meaning that the solution cannot be trusted. [We note that, in the case of the $2d$ Hubbard model, the series for $G_\mathrm{exact}$ also features a singularity at a real \textit{negative} $\xi_s$, which according to Eq.~(\ref{eq:ratio_test}) renders the exploding series coefficients sign-oscillating and in principle prevents the SDT from converging at all. However, this singularity is trivially eliminated by a conformal map at the start~\cite{Kim2021homotopy} and plays no role in the analysis of misleading convergence.]

In summary, we have found that misleading convergence is not always fatal for the self-consistent diagrammatic technique. There are regimes, such as that in the $2d$ Hubbard model with substantial doping, in which the diagrammatic technique finds the wrong answer because the exact solution is hidden beyond the convergence radius of its diagrammatic series. Since the singularity limiting the convergence radius in the real-positive complex half-plane is not due to a phase transition, it is separated from the real axis. This enables a modification of the self-consistent diagrammatic technique, whereby the partial sum of the series is replaced by its analytic continuation beyond the singularity. We have shown that the modified diagrammatic technique is at least as reliable as its textbook formulation in regimes where the letter obtains the correct answer, but also finds the exact solution when the original technique is plagued by the misleading convergence of the discovered nature. The indicator of the breakdown of the proposed approach, which constitutes the \textit{a priori} control of its accuracy, is a peculiar singularity structure of the obtained self-energy with $\xi_s$ on the real axis and $0 <\xi_s <1$. 

This approach opens the way for numerically exact calculations in important strongly-correlated regimes. When the solution belongs to the NP branch of the LWF, it remains to be understood what physics is missed by the P branch, to which the diagrammatic technique is fundamentally (or currently) limited, and whether this limitation is truly fundamental. A compelling physical picture behind the P-to-NP switchover has recently been developed through the studies of two-body correlation functions~\cite{Chalupa2018, Chalupa2021, Adler2022}. The single-boson-exchange decomposition~\cite{Adler2022} of the (generalized) charge susceptibility reveals a direct link between the development of the local magnetic moment at half-filling and the necessity for the solution to switch to the NP branch via the divergence of the irreducible vertex function. This implies that the skeleton diagrammatic technique is likely prohibited from accessing, e.g., the development of the magnetic stripe phase in the $2d$ Hubbard model at strong coupling and small doping ($\langle n \rangle \gtrsim 0.8$)~\cite{Zheng2017stripe, Ido2018no_superconductivity, Qin2019absence_of_superconductivity}. However, we find that the highest density at which the technique remains applicable \textit{increases} with lowering the temperature, so that already at $T \sim 0.5t$, controlled results for $\langle n \rangle \lesssim 0.7$ could be obtained. Since the critical temperature $T_c$ for the onset of superconductivity in the $2d$ Hubbard model is projected to rise dramatically with $U$ at $\langle n \rangle \gtrsim 0.7$~\cite{Simkovic_superconductivity2021}, the skeleton diagrammatic technique powered by DiagMC might be the optimal method for reliably demonstrating the high-$T_c$ superconductivity.

\begin{acknowledgments}
We are grateful to S. Adler, P. Chalupa-Gantner, M. Reitner, G. Sangiovanni, A. Toschi, F. Werner, and P. Werner for fruitful discussions and exchange of ideas. A.J.K. acknowledges support from ERC Consolidator Grant No. 724103 and DGIST Start-up Fund Program  of  the  Ministry  of Science and ICT (2022100006). We are grateful for computational resources to Grand Challenging Project of Supercomuting Bigdata Center, DGIST.
\end{acknowledgments}

\bibliography{ref.bib}

\begin{thebibliography}{39}
\expandafter\ifx\csname natexlab\endcsname\relax\def\natexlab#1{#1}\fi
\expandafter\ifx\csname bibnamefont\endcsname\relax
  \def\bibnamefont#1{#1}\fi
\expandafter\ifx\csname bibfnamefont\endcsname\relax
  \def\bibfnamefont#1{#1}\fi
\expandafter\ifx\csname citenamefont\endcsname\relax
  \def\citenamefont#1{#1}\fi
\expandafter\ifx\csname url\endcsname\relax
  \def\url#1{\texttt{#1}}\fi
\expandafter\ifx\csname urlprefix\endcsname\relax\def\urlprefix{URL }\fi
\providecommand{\bibinfo}[2]{#2}
\providecommand{\eprint}[2][]{\url{#2}}

\bibitem[{\citenamefont{Abrikosov et~al.}(1975)\citenamefont{Abrikosov,
  Gor'kov, and Dzyaloshinski}}]{AGD}
\bibinfo{author}{\bibfnamefont{A.~A.} \bibnamefont{Abrikosov}},
  \bibinfo{author}{\bibfnamefont{L.~P.} \bibnamefont{Gor'kov}},
  \bibnamefont{and} \bibinfo{author}{\bibfnamefont{I.~E.}
  \bibnamefont{Dzyaloshinski}}, \emph{\bibinfo{title}{Methods of Quantum Field
  Theory in Statistical Physics}} (\bibinfo{publisher}{Dover Publications
  Inc.}, \bibinfo{year}{1975}).

\bibitem[{\citenamefont{Georges et~al.}(1996)\citenamefont{Georges, Kotliar,
  Krauth, and Rozenberg}}]{Georges:1996zz}
\bibinfo{author}{\bibfnamefont{A.}~\bibnamefont{Georges}},
  \bibinfo{author}{\bibfnamefont{G.}~\bibnamefont{Kotliar}},
  \bibinfo{author}{\bibfnamefont{W.}~\bibnamefont{Krauth}}, \bibnamefont{and}
  \bibinfo{author}{\bibfnamefont{M.~J.} \bibnamefont{Rozenberg}},
  \bibinfo{journal}{Rev. Mod. Phys.} \textbf{\bibinfo{volume}{68}},
  \bibinfo{pages}{13} (\bibinfo{year}{1996}),
  \urlprefix\url{http://link.aps.org/doi/10.1103/RevModPhys.68.13}.

\bibitem[{\citenamefont{Maier et~al.}(2005)\citenamefont{Maier, Jarrell,
  Pruschke, and Hettler}}]{Maier2005}
\bibinfo{author}{\bibfnamefont{T.}~\bibnamefont{Maier}},
  \bibinfo{author}{\bibfnamefont{M.}~\bibnamefont{Jarrell}},
  \bibinfo{author}{\bibfnamefont{T.}~\bibnamefont{Pruschke}}, \bibnamefont{and}
  \bibinfo{author}{\bibfnamefont{M.~H.} \bibnamefont{Hettler}},
  \bibinfo{journal}{Rev. Mod. Phys.} \textbf{\bibinfo{volume}{77}},
  \bibinfo{pages}{1027} (\bibinfo{year}{2005}),
  \urlprefix\url{https://link.aps.org/doi/10.1103/RevModPhys.77.1027}.

\bibitem[{\citenamefont{Rohringer et~al.}(2018)\citenamefont{Rohringer,
  Hafermann, Toschi, Katanin, Antipov, Katsnelson, Lichtenstein, Rubtsov, and
  Held}}]{Rohringer2018}
\bibinfo{author}{\bibfnamefont{G.}~\bibnamefont{Rohringer}},
  \bibinfo{author}{\bibfnamefont{H.}~\bibnamefont{Hafermann}},
  \bibinfo{author}{\bibfnamefont{A.}~\bibnamefont{Toschi}},
  \bibinfo{author}{\bibfnamefont{A.~A.} \bibnamefont{Katanin}},
  \bibinfo{author}{\bibfnamefont{A.~E.} \bibnamefont{Antipov}},
  \bibinfo{author}{\bibfnamefont{M.~I.} \bibnamefont{Katsnelson}},
  \bibinfo{author}{\bibfnamefont{A.~I.} \bibnamefont{Lichtenstein}},
  \bibinfo{author}{\bibfnamefont{A.~N.} \bibnamefont{Rubtsov}},
  \bibnamefont{and} \bibinfo{author}{\bibfnamefont{K.}~\bibnamefont{Held}},
  \bibinfo{journal}{Rev. Mod. Phys.} \textbf{\bibinfo{volume}{90}},
  \bibinfo{pages}{025003} (\bibinfo{year}{2018}),
  \urlprefix\url{https://link.aps.org/doi/10.1103/RevModPhys.90.025003}.

\bibitem[{\citenamefont{Gull et~al.}(2011)\citenamefont{Gull, Millis,
  Lichtenstein, Rubtsov, Troyer, and Werner}}]{Gull2011}
\bibinfo{author}{\bibfnamefont{E.}~\bibnamefont{Gull}},
  \bibinfo{author}{\bibfnamefont{A.~J.} \bibnamefont{Millis}},
  \bibinfo{author}{\bibfnamefont{A.~I.} \bibnamefont{Lichtenstein}},
  \bibinfo{author}{\bibfnamefont{A.~N.} \bibnamefont{Rubtsov}},
  \bibinfo{author}{\bibfnamefont{M.}~\bibnamefont{Troyer}}, \bibnamefont{and}
  \bibinfo{author}{\bibfnamefont{P.}~\bibnamefont{Werner}},
  \bibinfo{journal}{Rev. Mod. Phys.} \textbf{\bibinfo{volume}{83}},
  \bibinfo{pages}{349} (\bibinfo{year}{2011}),
  \urlprefix\url{https://link.aps.org/doi/10.1103/RevModPhys.83.349}.

\bibitem[{\citenamefont{Blankenbecler et~al.}(1981)\citenamefont{Blankenbecler,
  Scalapino, and Sugar}}]{Blankenbecler1981}
\bibinfo{author}{\bibfnamefont{R.}~\bibnamefont{Blankenbecler}},
  \bibinfo{author}{\bibfnamefont{D.~J.} \bibnamefont{Scalapino}},
  \bibnamefont{and} \bibinfo{author}{\bibfnamefont{R.~L.} \bibnamefont{Sugar}},
  \bibinfo{journal}{Phys. Rev. D} \textbf{\bibinfo{volume}{24}},
  \bibinfo{pages}{2278} (\bibinfo{year}{1981}),
  \urlprefix\url{https://link.aps.org/doi/10.1103/PhysRevD.24.2278}.

\bibitem[{\citenamefont{Zhang and Krakauer}(2003)}]{Zhang2003}
\bibinfo{author}{\bibfnamefont{S.}~\bibnamefont{Zhang}} \bibnamefont{and}
  \bibinfo{author}{\bibfnamefont{H.}~\bibnamefont{Krakauer}},
  \bibinfo{journal}{Phys. Rev. Lett.} \textbf{\bibinfo{volume}{90}},
  \bibinfo{pages}{136401} (\bibinfo{year}{2003}),
  \urlprefix\url{https://link.aps.org/doi/10.1103/PhysRevLett.90.136401}.

\bibitem[{\citenamefont{Prokof'ev and Svistunov}(1998)}]{Prokofev1998}
\bibinfo{author}{\bibfnamefont{N.~V.} \bibnamefont{Prokof'ev}}
  \bibnamefont{and} \bibinfo{author}{\bibfnamefont{B.~V.}
  \bibnamefont{Svistunov}}, \bibinfo{journal}{Phys. Rev. Lett.}
  \textbf{\bibinfo{volume}{81}}, \bibinfo{pages}{2514} (\bibinfo{year}{1998}),
  \urlprefix\url{https://link.aps.org/doi/10.1103/PhysRevLett.81.2514}.

\bibitem[{\citenamefont{Prokof'ev and Svistunov}(2007)}]{Prokofev2007}
\bibinfo{author}{\bibfnamefont{N.}~\bibnamefont{Prokof'ev}} \bibnamefont{and}
  \bibinfo{author}{\bibfnamefont{B.}~\bibnamefont{Svistunov}},
  \bibinfo{journal}{Phys. Rev. Lett.} \textbf{\bibinfo{volume}{99}},
  \bibinfo{pages}{250201} (\bibinfo{year}{2007}),
  \urlprefix\url{https://link.aps.org/doi/10.1103/PhysRevLett.99.250201}.

\bibitem[{\citenamefont{{Van Houcke} et~al.}(2010)\citenamefont{{Van Houcke},
  Kozik, Prokof'ev, and Svistunov}}]{VanHoucke2010}
\bibinfo{author}{\bibfnamefont{K.}~\bibnamefont{{Van Houcke}}},
  \bibinfo{author}{\bibfnamefont{E.}~\bibnamefont{Kozik}},
  \bibinfo{author}{\bibfnamefont{N.}~\bibnamefont{Prokof'ev}},
  \bibnamefont{and}
  \bibinfo{author}{\bibfnamefont{B.}~\bibnamefont{Svistunov}},
  \bibinfo{journal}{Phys. Procedia} \textbf{\bibinfo{volume}{6}},
  \bibinfo{pages}{95} (\bibinfo{year}{2010}),
  \urlprefix\url{https://www.sciencedirect.com/science/article/pii/S1875389210006498}.

\bibitem[{\citenamefont{Kozik et~al.}(2010)\citenamefont{Kozik, {Van Houcke},
  Gull, Pollet, Prokof'ev, Svistunov, and Troyer}}]{Kozik2010}
\bibinfo{author}{\bibfnamefont{E.}~\bibnamefont{Kozik}},
  \bibinfo{author}{\bibfnamefont{K.}~\bibnamefont{{Van Houcke}}},
  \bibinfo{author}{\bibfnamefont{E.}~\bibnamefont{Gull}},
  \bibinfo{author}{\bibfnamefont{L.}~\bibnamefont{Pollet}},
  \bibinfo{author}{\bibfnamefont{N.}~\bibnamefont{Prokof'ev}},
  \bibinfo{author}{\bibfnamefont{B.}~\bibnamefont{Svistunov}},
  \bibnamefont{and} \bibinfo{author}{\bibfnamefont{M.}~\bibnamefont{Troyer}},
  \bibinfo{journal}{Europhys. Lett.} \textbf{\bibinfo{volume}{90}},
  \bibinfo{pages}{10004} (\bibinfo{year}{2010}),
  \urlprefix\url{https://doi.org/10.1209/0295-5075/90/10004}.

\bibitem[{\citenamefont{{Van Houcke} et~al.}(2012)\citenamefont{{Van Houcke},
  Werner, Kozik, Prokof'Ev, Svistunov, Ku, Sommer, Cheuk, Schirotzek, and
  Zwierlein}}]{VanHoucke2012}
\bibinfo{author}{\bibfnamefont{K.}~\bibnamefont{{Van Houcke}}},
  \bibinfo{author}{\bibfnamefont{F.}~\bibnamefont{Werner}},
  \bibinfo{author}{\bibfnamefont{E.}~\bibnamefont{Kozik}},
  \bibinfo{author}{\bibfnamefont{N.}~\bibnamefont{Prokof'Ev}},
  \bibinfo{author}{\bibfnamefont{B.}~\bibnamefont{Svistunov}},
  \bibinfo{author}{\bibfnamefont{M.~J.} \bibnamefont{Ku}},
  \bibinfo{author}{\bibfnamefont{A.~T.} \bibnamefont{Sommer}},
  \bibinfo{author}{\bibfnamefont{L.~W.} \bibnamefont{Cheuk}},
  \bibinfo{author}{\bibfnamefont{A.}~\bibnamefont{Schirotzek}},
  \bibnamefont{and} \bibinfo{author}{\bibfnamefont{M.~W.}
  \bibnamefont{Zwierlein}}, \bibinfo{journal}{Nat. Phys.}
  \textbf{\bibinfo{volume}{8}}, \bibinfo{pages}{366} (\bibinfo{year}{2012}),
  \urlprefix\url{http://dx.doi.org/10.1038/nphys2273
  papers3://publication/doi/10.1038/nphys2273}.

\bibitem[{\citenamefont{Kim et~al.}(2021)\citenamefont{Kim, Prokof'ev,
  Svistunov, and Kozik}}]{Kim2021homotopy}
\bibinfo{author}{\bibfnamefont{A.~J.} \bibnamefont{Kim}},
  \bibinfo{author}{\bibfnamefont{N.~V.} \bibnamefont{Prokof'ev}},
  \bibinfo{author}{\bibfnamefont{B.~V.} \bibnamefont{Svistunov}},
  \bibnamefont{and} \bibinfo{author}{\bibfnamefont{E.}~\bibnamefont{Kozik}},
  \bibinfo{journal}{Phys. Rev. Lett.} \textbf{\bibinfo{volume}{126}},
  \bibinfo{pages}{257001} (\bibinfo{year}{2021}),
  \urlprefix\url{https://link.aps.org/doi/10.1103/PhysRevLett.126.257001}.

\bibitem[{\citenamefont{Baym and Kadanoff}(1961)}]{Baym_Kadanoff}
\bibinfo{author}{\bibfnamefont{G.}~\bibnamefont{Baym}} \bibnamefont{and}
  \bibinfo{author}{\bibfnamefont{L.~P.} \bibnamefont{Kadanoff}},
  \bibinfo{journal}{Phys. Rev.} \textbf{\bibinfo{volume}{124}},
  \bibinfo{pages}{287} (\bibinfo{year}{1961}),
  \urlprefix\url{http://link.aps.org/doi/10.1103/PhysRev.124.287}.

\bibitem[{\citenamefont{Kozik et~al.}(2015)\citenamefont{Kozik, Ferrero, and
  Georges}}]{Kozik2015}
\bibinfo{author}{\bibfnamefont{E.}~\bibnamefont{Kozik}},
  \bibinfo{author}{\bibfnamefont{M.}~\bibnamefont{Ferrero}}, \bibnamefont{and}
  \bibinfo{author}{\bibfnamefont{A.}~\bibnamefont{Georges}},
  \bibinfo{journal}{Phys. Rev. Lett.} \textbf{\bibinfo{volume}{114}},
  \bibinfo{pages}{156402} (\bibinfo{year}{2015}),
  \urlprefix\url{https://link.aps.org/doi/10.1103/PhysRevLett.114.156402}.

\bibitem[{\citenamefont{Luttinger and Ward}(1960)}]{Luttinger1960}
\bibinfo{author}{\bibfnamefont{J.~M.} \bibnamefont{Luttinger}}
  \bibnamefont{and} \bibinfo{author}{\bibfnamefont{J.~C.} \bibnamefont{Ward}},
  \bibinfo{journal}{Phys. Rev.} \textbf{\bibinfo{volume}{118}},
  \bibinfo{pages}{1417} (\bibinfo{year}{1960}),
  \urlprefix\url{https://link.aps.org/doi/10.1103/PhysRev.118.1417}.

\bibitem[{\citenamefont{Sch{\"{a}}fer et~al.}(2013)\citenamefont{Sch{\"{a}}fer,
  Rohringer, Gunnarsson, Ciuchi, Sangiovanni, and Toschi}}]{Schafer2013}
\bibinfo{author}{\bibfnamefont{T.}~\bibnamefont{Sch{\"{a}}fer}},
  \bibinfo{author}{\bibfnamefont{G.}~\bibnamefont{Rohringer}},
  \bibinfo{author}{\bibfnamefont{O.}~\bibnamefont{Gunnarsson}},
  \bibinfo{author}{\bibfnamefont{S.}~\bibnamefont{Ciuchi}},
  \bibinfo{author}{\bibfnamefont{G.}~\bibnamefont{Sangiovanni}},
  \bibnamefont{and} \bibinfo{author}{\bibfnamefont{A.}~\bibnamefont{Toschi}},
  \bibinfo{journal}{Phys. Rev. Lett.} \textbf{\bibinfo{volume}{110}},
  \bibinfo{pages}{246405} (\bibinfo{year}{2013}),
  \urlprefix\url{https://link.aps.org/doi/10.1103/PhysRevLett.110.246405
  papers3://publication/doi/10.1103/PhysRevLett.110.246405}.

\bibitem[{\citenamefont{Rossi and Werner}(2015)}]{Rossi:2015cx}
\bibinfo{author}{\bibfnamefont{R.}~\bibnamefont{Rossi}} \bibnamefont{and}
  \bibinfo{author}{\bibfnamefont{F.}~\bibnamefont{Werner}},
  \bibinfo{journal}{J. Phys. A} \textbf{\bibinfo{volume}{48}},
  \bibinfo{pages}{485202} (\bibinfo{year}{2015}),
  \urlprefix\url{http://stacks.iop.org/1751-8121/48/i=48/a=485202?key=crossref.1bbe18fa768e2f6df9d10fbe499aed7a}.

\bibitem[{\citenamefont{Gunnarsson et~al.}(2017)\citenamefont{Gunnarsson,
  Rohringer, Sch{\"a}fer, Sangiovanni, and Toschi}}]{Gunnarsson:2017c}
\bibinfo{author}{\bibfnamefont{O.}~\bibnamefont{Gunnarsson}},
  \bibinfo{author}{\bibfnamefont{G.}~\bibnamefont{Rohringer}},
  \bibinfo{author}{\bibfnamefont{T.}~\bibnamefont{Sch{\"a}fer}},
  \bibinfo{author}{\bibfnamefont{G.}~\bibnamefont{Sangiovanni}},
  \bibnamefont{and} \bibinfo{author}{\bibfnamefont{A.}~\bibnamefont{Toschi}},
  \bibinfo{journal}{Phys. Rev. Lett.} \textbf{\bibinfo{volume}{119}},
  \bibinfo{pages}{056402} (\bibinfo{year}{2017}),
  \urlprefix\url{http://link.aps.org/doi/10.1103/PhysRevLett.119.056402}.

\bibitem[{\citenamefont{Tarantino et~al.}(2017)\citenamefont{Tarantino,
  Romaniello, Berger, and Reining}}]{Tarantino:2017dr}
\bibinfo{author}{\bibfnamefont{W.}~\bibnamefont{Tarantino}},
  \bibinfo{author}{\bibfnamefont{P.}~\bibnamefont{Romaniello}},
  \bibinfo{author}{\bibfnamefont{J.~A.} \bibnamefont{Berger}},
  \bibnamefont{and} \bibinfo{author}{\bibfnamefont{L.}~\bibnamefont{Reining}},
  \bibinfo{journal}{Phys. Rev. B} \textbf{\bibinfo{volume}{96}},
  \bibinfo{pages}{045124} (\bibinfo{year}{2017}),
  \urlprefix\url{http://link.aps.org/doi/10.1103/PhysRevB.96.045124}.

\bibitem[{\citenamefont{Thunström et~al.}(2018)\citenamefont{Thunström,
  Gunnarsson, Ciuchi, and Rohringer}}]{Thunstroem2018}
\bibinfo{author}{\bibfnamefont{P.}~\bibnamefont{Thunström}},
  \bibinfo{author}{\bibfnamefont{O.}~\bibnamefont{Gunnarsson}},
  \bibinfo{author}{\bibfnamefont{S.}~\bibnamefont{Ciuchi}}, \bibnamefont{and}
  \bibinfo{author}{\bibfnamefont{G.}~\bibnamefont{Rohringer}},
  \bibinfo{journal}{Phys. Rev. B} \textbf{\bibinfo{volume}{98}},
  \bibinfo{pages}{767} (\bibinfo{year}{2018}),
  \urlprefix\url{https://link.aps.org/doi/10.1103/PhysRevB.98.235107}.

\bibitem[{\citenamefont{Chalupa et~al.}(2018)\citenamefont{Chalupa, Gunacker,
  Sch{\"{a}}fer, Held, and Toschi}}]{Chalupa2018}
\bibinfo{author}{\bibfnamefont{P.}~\bibnamefont{Chalupa}},
  \bibinfo{author}{\bibfnamefont{P.}~\bibnamefont{Gunacker}},
  \bibinfo{author}{\bibfnamefont{T.}~\bibnamefont{Sch{\"{a}}fer}},
  \bibinfo{author}{\bibfnamefont{K.}~\bibnamefont{Held}}, \bibnamefont{and}
  \bibinfo{author}{\bibfnamefont{A.}~\bibnamefont{Toschi}},
  \bibinfo{journal}{Phys. Rev. B} \textbf{\bibinfo{volume}{97}},
  \bibinfo{pages}{245136} (\bibinfo{year}{2018}),
  \urlprefix\url{https://link.aps.org/doi/10.1103/PhysRevB.97.245136}.

\bibitem[{\citenamefont{Kim and Sacksteder}(2020)}]{KimReplica}
\bibinfo{author}{\bibfnamefont{A.~J.} \bibnamefont{Kim}} \bibnamefont{and}
  \bibinfo{author}{\bibfnamefont{V.}~\bibnamefont{Sacksteder}},
  \bibinfo{journal}{Phys. Rev. B} \textbf{\bibinfo{volume}{101}},
  \bibinfo{pages}{115146} (\bibinfo{year}{2020}),
  \urlprefix\url{https://link.aps.org/doi/10.1103/PhysRevB.101.115146}.

\bibitem[{\citenamefont{Chalupa et~al.}(2021)\citenamefont{Chalupa, Sch\"afer,
  Reitner, Springer, Andergassen, and Toschi}}]{Chalupa2021}
\bibinfo{author}{\bibfnamefont{P.}~\bibnamefont{Chalupa}},
  \bibinfo{author}{\bibfnamefont{T.}~\bibnamefont{Sch\"afer}},
  \bibinfo{author}{\bibfnamefont{M.}~\bibnamefont{Reitner}},
  \bibinfo{author}{\bibfnamefont{D.}~\bibnamefont{Springer}},
  \bibinfo{author}{\bibfnamefont{S.}~\bibnamefont{Andergassen}},
  \bibnamefont{and} \bibinfo{author}{\bibfnamefont{A.}~\bibnamefont{Toschi}},
  \bibinfo{journal}{Phys. Rev. Lett.} \textbf{\bibinfo{volume}{126}},
  \bibinfo{pages}{056403} (\bibinfo{year}{2021}),
  \urlprefix\url{https://link.aps.org/doi/10.1103/PhysRevLett.126.056403}.

\bibitem[{\citenamefont{{Van Houcke} et~al.}(2021)\citenamefont{{Van Houcke},
  {Kozik}, {Rossi}, {Deng}, and {Werner}}}]{VanHoucke2021}
\bibinfo{author}{\bibfnamefont{K.}~\bibnamefont{{Van Houcke}}},
  \bibinfo{author}{\bibfnamefont{E.}~\bibnamefont{{Kozik}}},
  \bibinfo{author}{\bibfnamefont{R.}~\bibnamefont{{Rossi}}},
  \bibinfo{author}{\bibfnamefont{Y.}~\bibnamefont{{Deng}}}, \bibnamefont{and}
  \bibinfo{author}{\bibfnamefont{F.}~\bibnamefont{{Werner}}},
  \bibinfo{journal}{arXiv e-prints} \bibinfo{eid}{arXiv:2102.04508}
  (\bibinfo{year}{2021}), \eprint{2102.04508}.

\bibitem[{\citenamefont{{Kim} et~al.}(2020)\citenamefont{{Kim}, {Werner}, and
  {Kozik}}}]{KimStrangeMetal}
\bibinfo{author}{\bibfnamefont{A.~J.} \bibnamefont{{Kim}}},
  \bibinfo{author}{\bibfnamefont{P.}~\bibnamefont{{Werner}}}, \bibnamefont{and}
  \bibinfo{author}{\bibfnamefont{E.}~\bibnamefont{{Kozik}}},
  \bibinfo{journal}{arXiv e-prints} \bibinfo{eid}{arXiv:2012.06159}
  (\bibinfo{year}{2020}), \eprint{2012.06159}.

\bibitem[{\citenamefont{{Adler} et~al.}(2022)\citenamefont{{Adler}, {Krien},
  {Chalupa-Gantner}, {Sangiovanni}, and {Toschi}}}]{Adler2022}
\bibinfo{author}{\bibfnamefont{S.}~\bibnamefont{{Adler}}},
  \bibinfo{author}{\bibfnamefont{F.}~\bibnamefont{{Krien}}},
  \bibinfo{author}{\bibfnamefont{P.}~\bibnamefont{{Chalupa-Gantner}}},
  \bibinfo{author}{\bibfnamefont{G.}~\bibnamefont{{Sangiovanni}}},
  \bibnamefont{and} \bibinfo{author}{\bibfnamefont{A.}~\bibnamefont{{Toschi}}},
  \bibinfo{journal}{arXiv e-prints} \bibinfo{eid}{arXiv:2212.09693}
  (\bibinfo{year}{2022}), \eprint{2212.09693}.

\bibitem[{\citenamefont{Prokof'ev and Svistunov}(2008)}]{Prokofev2008}
\bibinfo{author}{\bibfnamefont{N.~V.} \bibnamefont{Prokof'ev}}
  \bibnamefont{and} \bibinfo{author}{\bibfnamefont{B.~V.}
  \bibnamefont{Svistunov}}, \bibinfo{journal}{Phys. Rev. B}
  \textbf{\bibinfo{volume}{77}}, \bibinfo{pages}{125101}
  (\bibinfo{year}{2008}),
  \urlprefix\url{https://link.aps.org/doi/10.1103/PhysRevB.77.125101}.

\bibitem[{\citenamefont{Rossi et~al.}(2016)\citenamefont{Rossi, Werner,
  Prokof'ev, and Svistunov}}]{Rossi_shifted_action_2016}
\bibinfo{author}{\bibfnamefont{R.}~\bibnamefont{Rossi}},
  \bibinfo{author}{\bibfnamefont{F.}~\bibnamefont{Werner}},
  \bibinfo{author}{\bibfnamefont{N.}~\bibnamefont{Prokof'ev}},
  \bibnamefont{and}
  \bibinfo{author}{\bibfnamefont{B.}~\bibnamefont{Svistunov}},
  \bibinfo{journal}{Phys. Rev. B} \textbf{\bibinfo{volume}{93}},
  \bibinfo{pages}{161102} (\bibinfo{year}{2016}),
  \urlprefix\url{https://link.aps.org/doi/10.1103/PhysRevB.93.161102}.

\bibitem[{\citenamefont{Janke}(1998)}]{janke1998resummation}
\bibinfo{author}{\bibfnamefont{W.}~\bibnamefont{Janke}},
  \emph{\bibinfo{title}{Resummation of Divergent Perturbation Series:
  Introduction to Theory \& Guide to Practical Applications}}
  (\bibinfo{publisher}{World Scientific}, \bibinfo{year}{1998}).

\bibitem[{\citenamefont{Baker}(1961)}]{baker1961Dlog}
\bibinfo{author}{\bibfnamefont{G.~A.} \bibnamefont{Baker}},
  \bibinfo{journal}{Phys. Rev.} \textbf{\bibinfo{volume}{124}},
  \bibinfo{pages}{768} (\bibinfo{year}{1961}),
  \urlprefix\url{https://link.aps.org/doi/10.1103/PhysRev.124.768}.

\bibitem[{\citenamefont{Hunter and Baker}(1979)}]{Hunter1979}
\bibinfo{author}{\bibfnamefont{D.~L.} \bibnamefont{Hunter}} \bibnamefont{and}
  \bibinfo{author}{\bibfnamefont{G.~A.} \bibnamefont{Baker}},
  \bibinfo{journal}{Phys. Rev. B} \textbf{\bibinfo{volume}{19}},
  \bibinfo{pages}{3808} (\bibinfo{year}{1979}),
  \urlprefix\url{https://link.aps.org/doi/10.1103/PhysRevB.19.3808}.

\bibitem[{\citenamefont{\ifmmode~\check{S}\else \v{S}\fi{}imkovic and
  Kozik}(2019)}]{Simkovic2019}
\bibinfo{author}{\bibfnamefont{F.}~\bibnamefont{\ifmmode~\check{S}\else
  \v{S}\fi{}imkovic}} \bibnamefont{and}
  \bibinfo{author}{\bibfnamefont{E.}~\bibnamefont{Kozik}},
  \bibinfo{journal}{Phys. Rev. B} \textbf{\bibinfo{volume}{100}},
  \bibinfo{pages}{121102} (\bibinfo{year}{2019}),
  \urlprefix\url{https://link.aps.org/doi/10.1103/PhysRevB.100.121102}.

\bibitem[{\citenamefont{\ifmmode~\check{S}\else \v{S}\fi{}imkovic
  et~al.}(2020)\citenamefont{\ifmmode~\check{S}\else \v{S}\fi{}imkovic,
  LeBlanc, Kim, Deng, Prokof'ev, Svistunov, and Kozik}}]{Fedor2020}
\bibinfo{author}{\bibfnamefont{F.}~\bibnamefont{\ifmmode~\check{S}\else
  \v{S}\fi{}imkovic}}, \bibinfo{author}{\bibfnamefont{J.~P.~F.}
  \bibnamefont{LeBlanc}}, \bibinfo{author}{\bibfnamefont{A.~J.}
  \bibnamefont{Kim}}, \bibinfo{author}{\bibfnamefont{Y.}~\bibnamefont{Deng}},
  \bibinfo{author}{\bibfnamefont{N.~V.} \bibnamefont{Prokof'ev}},
  \bibinfo{author}{\bibfnamefont{B.~V.} \bibnamefont{Svistunov}},
  \bibnamefont{and} \bibinfo{author}{\bibfnamefont{E.}~\bibnamefont{Kozik}},
  \bibinfo{journal}{Phys. Rev. Lett.} \textbf{\bibinfo{volume}{124}},
  \bibinfo{pages}{017003} (\bibinfo{year}{2020}),
  \urlprefix\url{https://link.aps.org/doi/10.1103/PhysRevLett.124.017003}.

\bibitem[{\citenamefont{Kim et~al.}(2020)\citenamefont{Kim, Simkovic, and
  Kozik}}]{Kim_PRL_2020}
\bibinfo{author}{\bibfnamefont{A.~J.} \bibnamefont{Kim}},
  \bibinfo{author}{\bibfnamefont{F.}~\bibnamefont{Simkovic}}, \bibnamefont{and}
  \bibinfo{author}{\bibfnamefont{E.}~\bibnamefont{Kozik}},
  \bibinfo{journal}{Phys. Rev. Lett.} \textbf{\bibinfo{volume}{124}},
  \bibinfo{pages}{117602} (\bibinfo{year}{2020}),
  \urlprefix\url{https://link.aps.org/doi/10.1103/PhysRevLett.124.117602}.

\bibitem[{\citenamefont{Zheng et~al.}(2017)\citenamefont{Zheng, Chung, Corboz,
  Ehlers, Qin, Noack, Shi, White, Zhang, and Chan}}]{Zheng2017stripe}
\bibinfo{author}{\bibfnamefont{B.-X.} \bibnamefont{Zheng}},
  \bibinfo{author}{\bibfnamefont{C.-M.} \bibnamefont{Chung}},
  \bibinfo{author}{\bibfnamefont{P.}~\bibnamefont{Corboz}},
  \bibinfo{author}{\bibfnamefont{G.}~\bibnamefont{Ehlers}},
  \bibinfo{author}{\bibfnamefont{M.-P.} \bibnamefont{Qin}},
  \bibinfo{author}{\bibfnamefont{R.~M.} \bibnamefont{Noack}},
  \bibinfo{author}{\bibfnamefont{H.}~\bibnamefont{Shi}},
  \bibinfo{author}{\bibfnamefont{S.~R.} \bibnamefont{White}},
  \bibinfo{author}{\bibfnamefont{S.}~\bibnamefont{Zhang}}, \bibnamefont{and}
  \bibinfo{author}{\bibfnamefont{G.~K.-L.} \bibnamefont{Chan}},
  \bibinfo{journal}{Science} \textbf{\bibinfo{volume}{358}},
  \bibinfo{pages}{1155} (\bibinfo{year}{2017}),
  \urlprefix\url{https://www.science.org/doi/abs/10.1126/science.aam7127}.

\bibitem[{\citenamefont{Ido et~al.}(2018)\citenamefont{Ido, Ohgoe, and
  Imada}}]{Ido2018no_superconductivity}
\bibinfo{author}{\bibfnamefont{K.}~\bibnamefont{Ido}},
  \bibinfo{author}{\bibfnamefont{T.}~\bibnamefont{Ohgoe}}, \bibnamefont{and}
  \bibinfo{author}{\bibfnamefont{M.}~\bibnamefont{Imada}},
  \bibinfo{journal}{Phys. Rev. B} \textbf{\bibinfo{volume}{97}},
  \bibinfo{pages}{045138} (\bibinfo{year}{2018}),
  \urlprefix\url{https://link.aps.org/doi/10.1103/PhysRevB.97.045138}.

\bibitem[{\citenamefont{Qin et~al.}(2020)\citenamefont{Qin, Chung, Shi, Vitali,
  Hubig, Schollw\"ock, White, and Zhang}}]{Qin2019absence_of_superconductivity}
\bibinfo{author}{\bibfnamefont{M.}~\bibnamefont{Qin}},
  \bibinfo{author}{\bibfnamefont{C.-M.} \bibnamefont{Chung}},
  \bibinfo{author}{\bibfnamefont{H.}~\bibnamefont{Shi}},
  \bibinfo{author}{\bibfnamefont{E.}~\bibnamefont{Vitali}},
  \bibinfo{author}{\bibfnamefont{C.}~\bibnamefont{Hubig}},
  \bibinfo{author}{\bibfnamefont{U.}~\bibnamefont{Schollw\"ock}},
  \bibinfo{author}{\bibfnamefont{S.~R.} \bibnamefont{White}}, \bibnamefont{and}
  \bibinfo{author}{\bibfnamefont{S.}~\bibnamefont{Zhang}}
  (\bibinfo{collaboration}{Simons Collaboration on the Many-Electron Problem}),
  \bibinfo{journal}{Phys. Rev. X} \textbf{\bibinfo{volume}{10}},
  \bibinfo{pages}{031016} (\bibinfo{year}{2020}),
  \urlprefix\url{https://link.aps.org/doi/10.1103/PhysRevX.10.031016}.

\bibitem[{\citenamefont{\ifmmode~\check{S}\else \v{S}\fi{}imkovic
  et~al.}(2021)\citenamefont{\ifmmode~\check{S}\else \v{S}\fi{}imkovic, Deng,
  and Kozik}}]{Simkovic_superconductivity2021}
\bibinfo{author}{\bibfnamefont{F.}~\bibnamefont{\ifmmode~\check{S}\else
  \v{S}\fi{}imkovic}}, \bibinfo{author}{\bibfnamefont{Y.}~\bibnamefont{Deng}},
  \bibnamefont{and} \bibinfo{author}{\bibfnamefont{E.}~\bibnamefont{Kozik}},
  \bibinfo{journal}{Phys. Rev. B} \textbf{\bibinfo{volume}{104}},
  \bibinfo{pages}{L020507} (\bibinfo{year}{2021}),
  \urlprefix\url{https://link.aps.org/doi/10.1103/PhysRevB.104.L020507}.

\end{thebibliography}
\end{document}